\documentclass[aps,prd,onecolumn,amssymb]{revtex4}
\usepackage{graphicx,bm,color}
\usepackage[english]{babel}
\usepackage{amsmath}
\usepackage{amsfonts}

\begin{document}

\newcommand{\be}{\begin{equation}}
\newcommand{\ee}{\end{equation}}
\newcommand{\bea}{\begin{eqnarray}}
\newcommand{\eea}{\end{eqnarray}}
\newcommand{\nn}{\nonumber \\}
\newcommand{\e}{\mathrm{e}}

\title{Confronting dark energy models mimicking $\Lambda$CDM epoch with observational constraints: future cosmological perturbations decay or future Rip?}

\author{Artyom~V.~Astashenok$^{1}$ and Sergei~D.~Odintsov$^{2,3,4,5}$}
\affiliation{$^1$ Baltic Federal University of I. Kant, Department of Theoretical Physics, 236041, 14, Nevsky st., Kaliningrad, Russia \\
$^2$Instituci\`{o} Catalana de Recerca i Estudis Avan\c{c}ats (ICREA), Barcelona, Spain \\
$^3$Institut de Ciencies de l'Espai (CSIC-IEEC), Campus UAB, Torre C5-Par-2a pl, E-08193 Bellaterra (Barcelona), Spain\\
$^4$Eurasian International Center for Theor.Physics, Eurasian National University, Astana 010008, Kazahstan \\
$^5$Tomsk State Pedagogical University, Tomsk, Russia \\}

\begin{abstract}
We confront dark energy models which are currently similar to $\Lambda$CDM theory with observational data which include the SNe data, matter density perturbations and baryon acoustic oscillations data. DE cosmology under consideration may evolve to   Big Rip, type II or type III future singularity, or to Little Rip or Pseudo-Rip universe.
It is shown that  matter perturbations data  define more precisely
the possible deviation from $\Lambda$CDM model than consideration of SNe data only.
The combined data analysis proves that DE models under consideration are as consistent as $\Lambda$CDM model. We demonstrate that growth of matter density perturbations may occur at sufficiently small background density but still before the possible disintegration of bound objects (like clusters of galaxies, galaxies, etc) in Big Rip, type III singularity, Little Rip or Pseudo-Rip universe.
This new effect may bring the future universe to chaotic state well before disintegration or Rip.
\end{abstract}

\maketitle

\section{Introduction}

The late-time cosmic acceleration which seems to be proved by astrophysical data (see refs.\cite{Riess,Perlmutter}) opened the door for a number of (often exotic) theoretical  models of so-called dark energy (DE) (for recent review, see \cite{Dark-6,Ca}).
The dark energy contributing of nearly 72\% of the total mass energy of the universe \cite{Kowalski} has quite unusual properties like
negative pressure and/or negative entropy, invisibility in the early universe,
non-coupling with baryonic matter and effective non-observability, etc.
The DE properties may vary in wide limits, what depends from the specific DE model under discussion.

The negative pressure leads to that the equation of state (EoS) parameter $w_\mathrm{D0}$  is
negative:
\be
w_\mathrm{0}=p_\mathrm{D0}/\rho_\mathrm{D0}<0\, ,
\ee
where $\rho_\mathrm{D}$ is dark fluid energy-density and $p_\mathrm{D}$ is the
pressure. The subscript $0$ denotes that related quantities are considered at present time.

The data of observations (supernova data, baryon acoustic oscillations, etc.) indicate in favor of  vacuum energy/$\Lambda$CDM cosmology ($w=-1$). However the ultimate resolution of question about nature of dark energy seems to be impossible due to sufficiently large uncertainties in the determination of the DE EoS parameter $w=-1.04^{+0.09}_{-0.10}$ \cite{PDP,Amman}.

The phantom model ($w<-1$) firstly proposed in ref.\cite{Caldwell} probably is the most exotic model of dark energy. This model does not contradict the cosmological tests based on present data although from theoretical point of view the phantom field is unstable \cite{Carrol} because  the violation of all
energy conditions occurs.

The simplest phantom energy model with $w=\mbox{const}$ leads to so-called Big Rip future singularity\cite{Caldwell,Frampton,S,BR,Nojiri}. The scale factor becomes infinite at a finite time. Another types of finite-time future singularities are type II singularity\cite{Barrow} where the second derivative of scale factor becomes infinite at finite time while the first derivative is finite and type III singularity where the first derivative of scale factor diverges.

If $w$ decreases sufficiently rapidly with increase of phantom energy density the so-called Little Rip or Pseudo-Rip may be realized \cite{Frampton-2,Frampton-3,Astashenok}. In the first case the acceleration of universe increases but the scale factor remains finite always. Of course, the disintegration of bound structures occurs as in a case of Big Rip cosmology. For Pseudo-Rip universe, dark fluid energy-density asymptotically tends to constant value, i.e. this phantom energy model mimics vacuum energy from some moment of time.

To describe dark energy fluid we use the equation of state (EoS):
\be
\label{EoS-0}
p_{D}=F(\rho_{D})\, ,
\ee
where $F$ is a function of the energy density. The evolution of the universe then depends on the choice of the EoS.

In this paper we confront   DE models mimicking currently $\Lambda$CDM cosmology  with the combined observational data including the luminosity distance  modulus vs redshift for SNe Ia, the baryon acoustic oscillations (BAO) and matter and DE density perturbations. We demonstrate that  the models under discussion are totally viable and not-distinguishable from $\Lambda$CDM, while their future evolution may vary in a number of ways. The paper is organized as follows. In section II we present brief overview of the EoS fluid formalism. In section III  the
main constraints from observational data  are discussed.
In section IV we confront DE models with Big Rip or type III future singularity as well as Little Rip model with SNe data, BAO data and matter perturbations. It is established the excellent coincidence with $\Lambda$CDM model predictions at current universe. The consideration of DE and matter perturbations shows that growth of the energy-density may occur at sufficiently small background density but still before the possible disintegration of bound objects in the Rip universe. This new effect may bring the future universe to the separation in the several domains with different values of energy-density, i.e. kind of chaotic universe may emerge. Section V is devoted to comparison of DE models of Pseudo-Rip type or with type II future singularity with combined data. The region of parameters where such models effectively coincide with current $\Lambda$CDM cosmology are established.
 In Conclusion section some outlook is given.
In the Appendix we briefly describe the properties of Quasi-Rip universe.

\section{Scalar dark energy models}

For the spatially-flat Friedmann-Robertson-Walker (FRW) universe with metric
\be
ds^{2}=dt^{2}-a^{2}(t)(dx^{2}+dy^{2}+dz^{2})\, ,
\ee
 the FRW equations are given by
\be
\label{Fried1}
\left(\frac{\dot a}{a}\right)^2 = \frac{\rho}{3}\, , \quad
\dot{\rho} = -3\left(\frac{\dot a}{a}\right)(\rho + p)\, ,
\ee
where $\rho$ and $p$ are the total energy density and pressure,
$a$ is the scale factor, $\dot{}=d/dt$,
and we use the natural system of units in which $8\pi G=c=1$. For simplicity we assume that universe is filled only by cold dark matter, baryon matter and dark energy i.e. $\rho=\rho_{D}+\rho_{m}$ and $p=p_{D}$.

One can rewrite the dark energy EoS (\ref{EoS-0}) in the following form:
\be \label{EoS}
p_\mathrm{D}=-\rho_\mathrm{D}-g(\rho_\mathrm{D})\, ,
\ee
where $g(\rho_\mathrm{D})$ is some function. The case $g(\rho_\mathrm{D})>0$
corresponds
to the EoS parameter $w<-1$ (phantom) while the case $g(\rho_{D})<0$ corresponds to
the EoS
parameter $w>-1$. Assuming that dark energy dominates, one can neglect the contribution
of other components (matter, dark matter). Then from Eq.~(\ref{Fried1}), one
can get the following expression for time variable:
\be
\label{trho}
t-t_{0} \approx \frac{1}{\sqrt{3}}\int^{\rho_{D}}_{\rho_{D0}} \frac{d \rho_{D}}{\rho^{1/2}_{D}g(\rho_{D})}.
\ee
For current time we choose $t_{0}=0$. The quintessence
energy-density decreases with time ($\rho_{D}<\rho_{D0}$), while the phantom energy-density increases ($\rho_{D}>\rho_{D0}$). For scale factor we have the following expression:
\be
\label{arho}
a = a_{0}\exp\left(\frac{1}{3}\int^{\rho_{D}}_{\rho_{D0}} \frac{d\rho_{D}}{g(\rho_{D})}\right)\, ,
\ee
For simplicity $a_{0}=1$ is chosen. A finite-time singularity occurs if the integral (\ref{trho}) converges at $\rho\rightarrow\infty$. The scale factor in this case may become infinite (Big Rip) or may remain finite but a singularity ($\rho\rightarrow\infty$) occurs. This is a Type III singularity \cite{Nojiri-2}. The type II singularity realizes if $g(\rho_{D})\rightarrow\pm \infty$ at
$\rho_{D}=\rho_{f}$. The pressure of the dark energy becomes infinite at a
finite energy density. The second derivative of the scale factor diverges while
the first derivative remains finite.

Non-singular evolution corresponds to the case when integral (\ref{trho}) diverges at $\rho_{D}\rightarrow\infty$ (Little Rip \cite{Frampton-2}, \cite{Astashenok,others}) or  (\ref{trho}) diverges at $\rho_{D}\rightarrow\rho_{f}$\cite{Frampton-2}.
In the last case the dark energy density asymptotically tends to a constant value (``effective
cosmological constant'' or pseudo-rip).

The so-called quasi-rip \cite{Wei-0}, \cite{Piao} corresponds to (inverse) crossing of phantom divide line (``(de)phantomization" \cite{Andr}). One can show that in terms of EoS this case corresponds to multiply-valued function $g(\rho_{D})$ and phase transition at some value of scale factor may occur.

\section{Observational data}

Confrontation of the theoretical models with observational data includes the comparison with several observational constraints:

(i) the luminosity distance moduli to type Ia supernovae from the Supernova Cosmology Project \cite{Amanullah},

(ii) BAO (see for example \cite{Eisenstein}),

(iii) the data for the growth factor for density perturbations (see, for instance,refs.\cite{Mukhanov}, \cite{Christopherson}) from Lyman-$\alpha$ forest in the Sloan Digital Sky Survey and galaxy redshift distortions (for instance,refs.\cite{Guzzo}, \cite{Colles}).

Let us consider these observational bounds in detail.

(i) The distance modulus for a supernova with redshift $z=a_{0}/a-1$ is
\be
\mu(z)=\mbox{const}+5\log D(z)\, ,
\ee
where $D(z)$ is the luminosity distance. As is well-known
\be
\label{DLSC}
D_{L}(z)=\frac{c}{H_{0}}(1+z)\int_{0}^{z}
h^{-1}(z)d z, \quad h^{2}(z)=\rho(z)/\rho_{0}.
\ee
Here $c$ is speed of light and $H_{0}$ is Hubble parameter. We used $H_0 =72$ km/s/Mpc from the Hubble Space Telescope key project \cite{HST}. The best fit for SNe is given in the framework of $\Lambda$CDM cosmology. For such a model (which we call the ``standard cosmology'' (SC)), one obtains
\be
h(z)=(\Omega_{m0}(1+z)^{3}+\Omega_{\Lambda0})^{1/2}
\ee
Here, $\Omega_{m0}$ is the fraction of the total density contributed by matter at present time, and $\Omega_{\Lambda}$ is the fraction contributed by the vacuum energy density. Following the approach of ref.\cite{Wei} one can exclude the SNe samples satisfying the condition $|\mu_{obs}-\mu_{SC}|/\sigma_{obs}>1.9$. Therefore in the data set of 580 SNe from \cite{Union2} we exclude the following 50 SNe samples:

1992bs, 1992bp, 1995ac, 1999bm,  1997o, 2001hu, 1998ba, 04Pat, 05Red, 03D4au,

04D3gt, 04D3cp, 03D4at, 03D1fc, 04D3co, 03D4dy, 04D3oe, 04D1ak, 03D1co, b010,

1995aq, f076, g050, k430, m138, 2006br, 2006cm, 2006cf, 2007ca, 2004gc,

10106, 2005ll, 2005lp, 2005fp, 2005gr, 2005ia, 1997aj, f308, e140, d084,

2002hu, 2002ju, 2003ch, 2005gs, 2005hv, 2005ig, 2005jj, 1997k, g120, 05Str.

(ii) In Ref. \cite{Eisenstein} it was suggested that instead of taking the position of an acoustic peak one should measure the large-scale correlation function at 100$h^{-1}$ Mpc separation using the sample of 46748 luminous red galaxies (LRG)selected from the SDSS (Sloan Digital Sky Survey) sample. The appropriate quantity to be measured is known as  A parameter and reads as
\be \label{BAO}
A\equiv D_{V}(z_{0})\frac{\Omega^{1/2}_{m0}H_{0}}{z_{0}c}
\ee
In Eq. (\ref{BAO}) the dilaton scale $D_{V}(z)$ is defined as
$$
D_{V}(z)=\left[\frac{D_{M}(z)^{2}cz}{H(z)}\right]^{1/3}
$$
where $D_{M}(z)$ is the comoving angular diameter distance. The redshift $z_{0}=0.35$.  We have for the parameter A the following relation
\be\label{BAO-2}
A=\sqrt{\Omega_{m0}}h(z_{0})^{-1/3}\left[\frac{1}{z_{0}}\int_{0}^{z_{0}}h^{-1}(z)dz\right]^{2/3}.
\ee

The observational value of BAO parameter is $A=0.469\pm0.017$.

(iii) As it was shown in ref.\cite{Christopherson} one can neglect density perturbations of dark energy. In this case the dark matter perturbations effectively decouple from DE perturbations. The equation that determines the evolution of the density contrast $\delta$ in a flat background filled by the
 matter with density $\rho_{m}$ is
\be\label{Perturb}
\ddot{\delta}_{m}+2H\dot{\delta}_{m}=\frac{1}{2}\rho_{m}\delta_{m}
\ee
It is convenient to introduce the function of growth rate of perturbations $f=d\ln\delta_{m}/d\ln a$. Using FRW equations one can get the following equation for $f$:
\be
\frac{df}{d\ln a}+f^{2}+\left(\frac{\dot{H}}{H^{2}}+2\right)f-\frac{3}{2}\Omega_{m}=0,
\ee
where $\Omega_{m}$ is matter fraction of the total energy-density: $\Omega_{m}=\Omega_{m0}(1+z)^{3}/h^{2}(z)$. Finally, using relation
$$
\frac{d}{d\ln a}=-(1+z)\frac{d}{dz}
$$
and taking into account that
$$
2\frac{\ddot{a}}{a}+\frac{\dot{a}^{2}}{a^{2}}=-p_{D}
$$
we get
\be \label{Perturb-2}
-(1+z)\frac{df}{dz}+f^{2}+\left(\frac{1}{2}+\frac{3}{2}\Omega_{D}+\frac{3}{2}\frac{g(\rho_{D})}{\rho}\right)f-\frac{3}{2}\Omega_{m}=0.
\ee
where $\Omega_{D}=\rho_{D}/\rho$. For dark fluid with given EoS one can find DE
density as function of redshift $z$. Then, Eq. (\ref{Perturb-2}) can be solved numerically. The observational data for growth function $f$ at various redshifts are given in table I.

\begin{table}
\label{Table1}
\begin{centering}
\begin{tabular}{|c|c|c|}
  \hline
  $z$ & $f_{obs}$ & Ref. \\
  \hline
  0.15 & $0.51\pm0.11$ & \cite{Hawkins}, \cite{Verde} \\
  0.35 & $0.70\pm0.18$ & \cite{Tegmark} \\
  0.55 & $0.75\pm0.18$ & \cite{Ross} \\
  1.4 & $0.90\pm0.24$ & \cite{Angela} \\
  3.0 & $1.46\pm0.29$ & \cite{Mcdonald} \\
  \hline
\end{tabular}
\caption{The available data growth function $f$ at various redshifts from the change of power spectrum Ly-$\alpha$
forest data in SDSS.}
\end{centering}
\end{table}

In the data analysis we use $\chi^{2}$ statistics. The $\chi^{2}$ value for some physical quantity $x$ is given by equation
\be \label{CHI}
\chi^{2}_{x}=\frac{(x_{th}-x_{obs})^{2}}{\sigma^{2}_{x}}
\ee
where $x_{th}$ is theoretically predicted value of $x$, $x_{th}$ is experimentally measured value and $\sigma_{x}$ is standard deviation. For the data set the total $\chi^{2}$ is the sum of all $\chi^{2}_{x}$.

\section{Dark Energy model I: Little-Rip, Big-Rip and type III future singularity}

Let us start from the simple model with following EoS:
\be\label{LR}
g(\rho_{D})=\alpha^{2}\rho_{D0}\left(\frac{\rho_{D}}{\rho_{D0}}\right)^{\beta}
\ee
where $\alpha$ and $\beta$ are dimensionless constants. If $\beta=1$ we have ordinary phantom energy model with constant EoS parameter $w=-1-\alpha^{2}$. From Eqs. (\ref{trho}), (\ref{arho})  one can see that for various $\beta$ the model (\ref{LR}) describes three types of future universe evolution:

(a) Little Rip if $\beta\leq 1/2$,

(b) Big Rip if $1/2<\beta\leq 1$ and

(c) type III singularity if $\beta>1$.

The time left before finite-time future singularity can be estimated as following
\be
t_{f}-t_{0}\approx \frac{2}{3^{1/2}(2\beta-1)}\frac{1}{\alpha^{2}\rho^{1/2}_{D0}}.
\ee
Note that $\rho_{D0}=\Omega_{D0}\rho_{0}=3\Omega_{D0}H^{2}_{0}$ and therefore
$$
t_{f}-t_{0}=\frac{2}{3\alpha^{2}(2\beta-1)\Omega^{1/2}_{D0}}\frac{1}{H_{0}}.
$$

Let us restrict ourselves to cases (b) and (c). The results for Little Rip models are similar. From Eq. (\ref{arho}) one can derive the dependence of dark energy from redshift:
\be \label{RHOZLR}
\rho_{D}(z)=\left\{\begin{array}{cc}\rho_{D0}\left\{1-3\alpha^{2}(1-\beta)\ln(1+z)\right\}^{\frac{1}{1-\beta}},\quad \beta\neq 1,\\
\rho_{D0}(1+z)^{-3\alpha^{2}},\quad \beta=1.
\end{array}\right.
\ee

\textbf{SNe data analysis only.} Using Eq. (\ref{RHOZLR}) one can calculate the theoretical dependence $\mu(z)$ from (\ref{DLSC}). Of course the best-fit of SNe data is $\Lambda$CDM-model ($\alpha=0$). For $\Omega_{D0}=0.72$  the value of $\chi^{2}_{SN}$ is minimal: $\chi^{2}_{SN}=347.06$. For $\alpha\neq 0$ and given $\beta$ one can select the parameter $\Omega_{D0}$ to describe the experimental data with good accuracy. In table II the optimal $\Omega_{D0}$ for some values of parameters $\beta$ and $\alpha$ are given. We also calculated the $\chi^{2}_{SN}$ for comparison with $\Lambda$CDM-model and the time left for future singularity $t_{f}$ (we restrict ourselves to case $\beta>1/2$).

\begin{table}
\label{Table2}
\begin{centering}

\textbf{DE models with type III singularity: analysis of SNe data}

\begin{tabular}{|c|c|c|c|c|c|c|c|c|c|}
  \hline
& \multicolumn{3}{c}{$\beta=1.5$}\vline & \multicolumn{3}{c}{$\beta=2.0$}\vline & \multicolumn{3}{c}{$\beta=3.0$}\vline\\
  \hline
  $\alpha^{2}$ & $\Omega_{D0}$ & $t_{f}$, Gyr & $\chi^{2}_{SN}$ & $\Omega_{D0}$ & $t_{f}$, Gyr & $\chi^{2}_{SN}$ & $\Omega_{D0}$ & $t_{f}$, Gyr & $\chi^{2}_{SN}$\\

\hline
  0.1 & 0.68 & 55.0 & 347.44 & 0.68 & 36.7 & 347.42 & 0.68 & 22.0 & 347.41 \\
  0.2 & 0.65 & 28.1 & 348.62 & 0.65 & 18.7 & 348.57 & 0.65 & 11.2 & 348.57 \\
  0.3 & 0.62 & 19.1 & 350.18 & 0.62 & 12.7 & 350.28 & 0.63 & 7.6 & 350.30 \\
  0.4 & 0.60 & 14.6 & 352.36 & 0.60 & 9.7 & 352.40 & 0.61 & 5.8 & 352.41 \\
  0.5 & 0.58 & 11.9 & 354.80 & 0.58 & 7.9 & 354.93 & 0.59 & 4.7 & 354.90 \\
  \hline
\end{tabular}

\vspace{16pt}

\textbf{DE models with Big Rip singularity: analysis of SNe data}

\begin{tabular}{|c|c|c|c|c|c|c|c|c|c|}
  \hline
& \multicolumn{3}{c}{$\beta=0.55$}\vline & \multicolumn{3}{c}{$\beta=0.75$}\vline & \multicolumn{3}{c}{$\beta=0.95$}\vline\\
  \hline
  $\alpha^{2}$ & $\Omega_{D0}$ & $t_{f}$, Gyr & $\chi^{2}_{SN}$ & $\Omega_{D0}$ & $t_{f}$, Gyr & $\chi^{2}_{SN}$ & $\Omega_{D0}$ & $t_{f}$, Gyr & $\chi^{2}_{SN}$\\

\hline
  0.1 & 0.68 & $1.01\times10^{3}$ & 347.48 & 0.68 & 219.90 & 347.47 & 0.68 & 123.07 & 347.46 \\
  0.2 & 0.64 & 566.67 & 348.41 & 0.64 & 112.5 & 348.45 & 0.64 & 62.5 & 348.51 \\
  0.3 & 0.61 & 386.96 & 349.91 & 0.61 & 76.8 & 350.04 & 0.61 & 42.7 & 350.19 \\
  0.4 & 0.59 & 295.09 & 352.03 & 0.59 & 59.0 & 352.01 & 0.59 & 32.8 & 352.09 \\
  0.5 & 0.56 & 242.31 & 354.06 & 0.56 & 48.0 & 354.48 & 0.57 & 26.7 & 354.49 \\
  \hline
\end{tabular}
\caption{The optimal value of the parameter $\Omega_{D0}$, time left before future singularity and corresponding value of $\chi^{2}$ for some $\alpha^{2}$ and $\beta$.}

\end{centering}

\end{table}

One can see that in principle the parameters of our model can vary in sufficiently wide limits. However, the consideration of data for matter perturbations allows to restrict the  range of these parameters.

\textbf{Analysis of matter density perturbations data.} Again the best-fit for observational data is $\Lambda$CDM-model. For $\Omega_{D0}=0.72$ we find that $\chi^{2}_{DP}+\chi^{2}_{SN}$ is minimal at $f(0)=0.503$.  The value of $\chi^{2}_{DP}$ is 0.879 (within 95\% C.L). The consideration of matter perturbations data only leads to minimal value $\chi^{2}_{DP}=0.34$ at $\Omega_{D0}=0.78$ and $f(0)=0.441$. However, in this case the $\chi^{2}_{SN}$ is sufficiently large: $\chi^{2}_{SN}=370.01$ although this lies within 95\% confidence level.

For given $\beta$ one can further vary parameter $\alpha$ and find the optimal values of $\Omega_{D0}$ for fitting SNe data. Then one can find $f(0)$ from approximation of matter perturbations data. Hence, we find the maximal value of $\alpha$ at which our model fits the matter perturbations data with 95\% C.L. The results of our calculation for the model with future Big Rip singularity are given in table III. For DE model with type III future singularity the results are similar (see also \cite{pol}). We find that maximal values of parameters $\alpha^{2}_{max}$, $\Omega_{D0}$ and $f(0)$ in fact do not depend from $\beta$. It is interesting to note that SNe data description for these models coincides  with $\Lambda$CDM cosmology up to excellent accuracy.

The current EoS parameter for the model under consideration is
$$
w_{0}=-1-\alpha^{2}.
$$

Therefore small values of parameter $\alpha^{2}$ correspond to small deviation of $w$ from $-1$.

\begin{table}
\label{Table3}

\begin{centering}

\textbf{DE models with Big Rip or type II future singularity: analysis of SNe+DP data}

\begin{tabular}{|c|c|c|c|c|c|c|}
\hline
$\beta$ & $0.55$ & $0.75$ & $0.95$ & $1.5$ & $2$ & $3$ \\
\hline
$\alpha^{2}_{max}$ & \multicolumn{6}{c}{0.03}\vline \\
\hline
$\Omega_{D0}$ & \multicolumn{6}{c}{0.71}\vline  \\
\hline
$f(0)$ & \multicolumn{6}{c}{0.514}\vline  \\
\hline
$t_{fmin}$, Gyr & $3.6\times10^{3}$ & 717.3 & 398.5 & 179.3 & 119.6 & 71.7\\
\hline
$\chi^{2}_{SN} $ & \multicolumn{6}{c}{347.28}\vline  \\
\hline
$A$ & \multicolumn{6}{c}{0.483}\vline  \\
\hline
\end{tabular}
\caption{The maximal value of parameter $\alpha^{2}$ (and correspondingly the minimal time before singularity) for given $\beta$ at which the model (\ref{LR}) describes the density perturbations data with 95 \% C.L. The optimal value of $\Omega_{D0}$ for fitting SNe data, $\chi^{2}_{SN}$ and BAO parameter are also given.}
\end{centering}
\end{table}

 The dependence of $\ln a$ as function of the time is depicted for some parameters (figure 1). For the illustration, on the next figure the comparison between observational data for density perturbations and theoretical predictions is made. There is no significant difference between standard cosmology and our models.

\begin{figure}
  \includegraphics[scale=1]{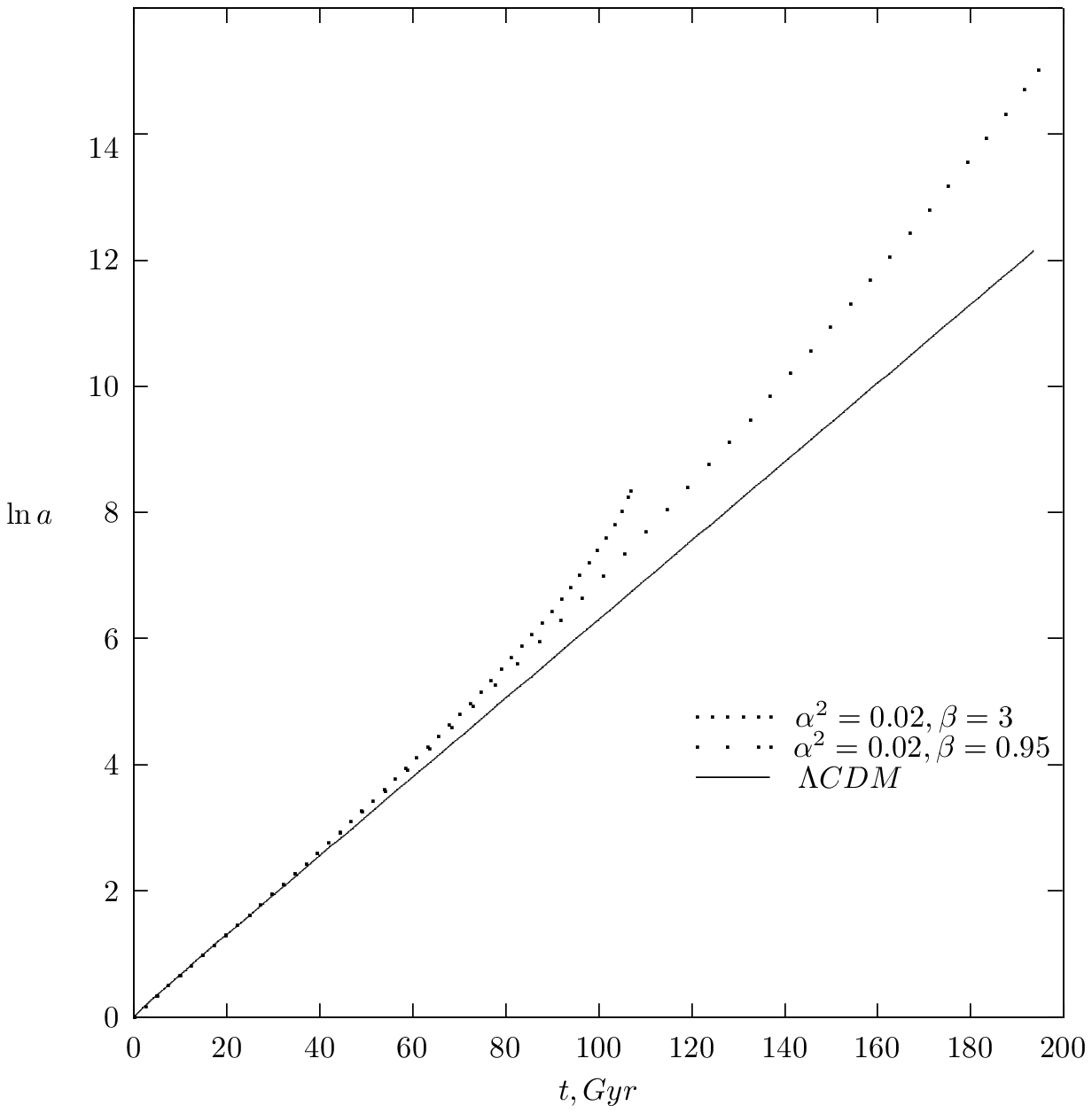}\\
  \caption{The dependence of scale factor from the time for model (\ref{LR}) with $\alpha^{2}=0.02$, $\beta=0.95$ (big rip) and $\beta=3$ (type III singularity). In the interval $\sim50$ Gyr the universe expansion  in these models is very close to the one  expected from $\Lambda$CDM cosmology.}
\end{figure}

\begin{figure}
  \includegraphics[scale=1]{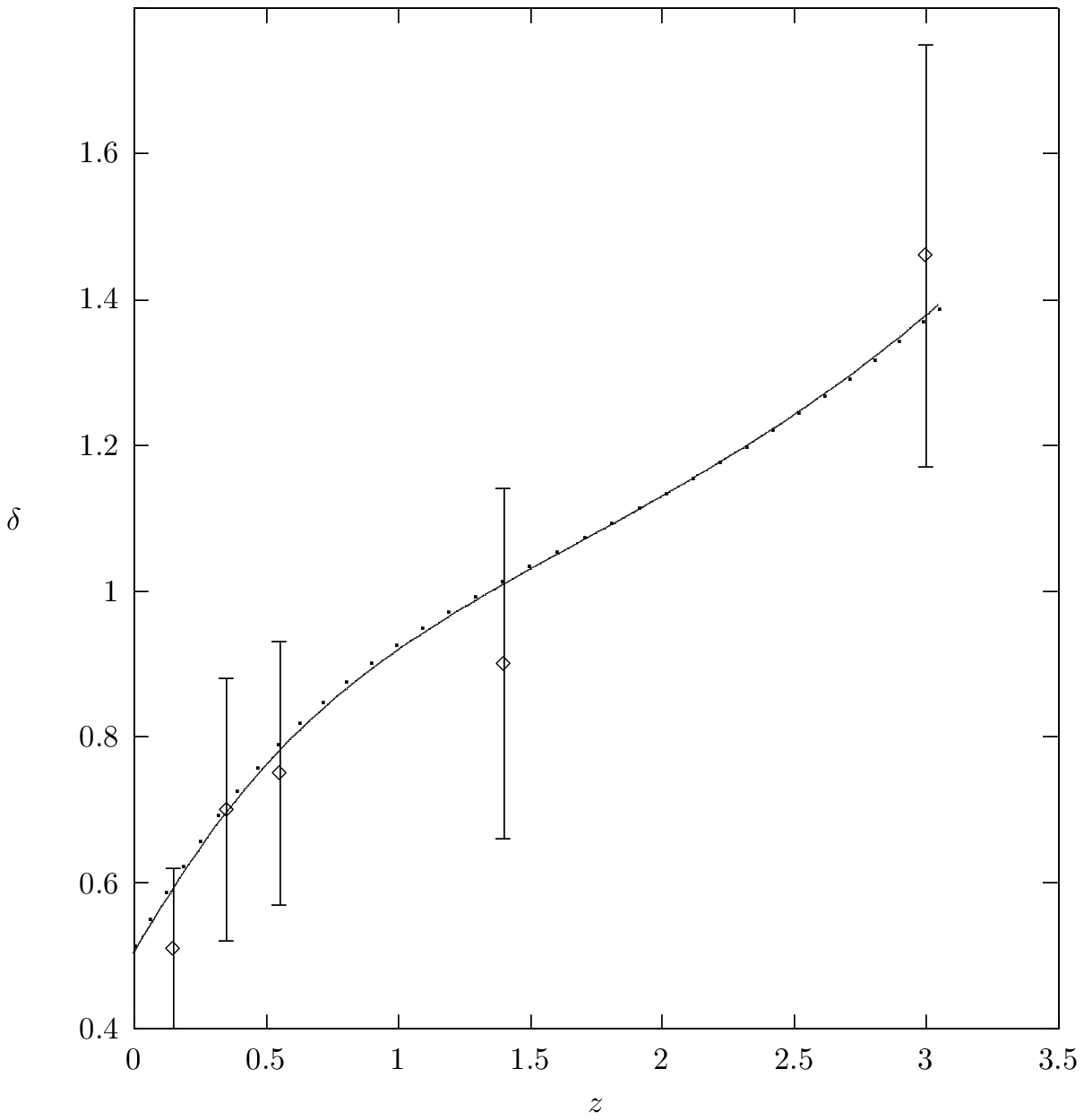}\\
  \caption{The comparison of observational data with theoretical predictions for the model (\ref{LR}) with $\alpha^{2}=0.02$, $\beta=0.95$ (Big Rip) and $\beta=3$ (type III singularity). One can see the coincidence with $\Lambda$CDM cosmology.}
\end{figure}

\textbf{Another DE model with type III singularity.} One can consider another DE model  which is more close to $\Lambda$CDM-cosmology. This model firstly was considered in \cite{Astashenok-2}, \cite{Astashenok-3}. From the following EoS
\be
\label{EoSBFS}
F(\rho)=-\beta^2 a_{f}^{\epsilon}\rho_{D}^{1+\epsilon/3}\, ,
\ee
where $\beta$, $a_{f}$, and $\epsilon$ are positive constants one can find the dependence of the dark energy density on the scale factor
\be
\rho_{D}=\beta^{-6/\epsilon}\left(a_\mathrm{f}^{\epsilon}-a^{\epsilon}
\right)^{-3/\epsilon}\, .
\ee

For dimensionless Hubble parameter as function of redshift we have therefore
\be
h^{2}(z)=\Omega_{m}(1+z)^{3}+\Omega_\mathrm{D}(1+z)^{3}\left(\frac{N_{0}-1}{N_{0}
(1+ z)^{\epsilon}-1}\right)^{3/\epsilon},\quad N_{0}=(a_{f}/a_{0})^{\epsilon}.
\ee

One can see that for large $N_{0}$ our model mimics $\Lambda$CDM cosmology with excellent precision. Therefore our model can fit the Supernova Cosmological Project data. For $N_{0}\gg 1$ the dark energy density is nearly constant in the
interval $0<t<t_{0}$, i.e. the model (\ref{EoSBFS}) mimics a cosmological
constant in the past but it leads to a finite-time future singularity.

The current EoS parameter is
$$
w_{0}=-\frac{N_{0}}{N_{0}-1}
$$

For given value of $w_{0}$ and $\epsilon$ one can find that such model describes the observational data with good accuracy. The results are given in table IV. One can see that model (\ref{EoSBFS}) is more close to $\Lambda$CDM cosmology than DE model with type III singularity considered above. The model describes supernova data with the same precision as the $\Lambda$CDM model although for density perturbations data the agreement is slightly worse than in the case of standard cosmological model. It is interesting to note that agreement between this DE model and observational data is better for large values of $\epsilon$. The model (\ref{EoSBFS}) coincides with $\Lambda$CDM model in past with excellent precision but its future evolution shows radically different dynamics
 (see fig. 3).

\begin{table}
\label{Table4}
\begin{tabular}{|c|c|c|c|c|c|c|c|c|c|c|}
\hline
& \multicolumn{3}{c}{$w_{0}=-1.10$}\vline & \multicolumn{3}{c}{$w_{0}=-1.05$}\vline & \multicolumn{3}{c}{$w_{0}=-1.02$}\vline \\
\hline
$\epsilon$ & 2 & 5 & 10  & 2 & 5 & 10 & 2 & 5 & 10\\
\hline
$\Omega_{D0}$ & \multicolumn{9}{c}{0.71}\vline \\
\hline
$f(0)$ & 0.514 & 0.514 & 0.514 & 0.514 & 0.513 & 0.514 & 0.513 & 0.513 & 0.513 \\
$\chi^{2}_{SN}$ & 350.97 & 349.04 & 347.88 & 347.68  & 347.29 & 347.08 & 346.98 & 346.97 & 346.98  \\
$\chi^{2}_{DP}$ & 1.26 & 1.20 & 1.15 & 1.15 & 1.13 & 1.11 & 1.10 & 1.08 & 1.07 \\
$t_{f}$, Gyr & 13.5 & 6.1 & 3.2 & 18.1 & 8.0 & 4.1 & 24.7 & 10.7 & 5.4 \\
\hline
\end{tabular}
\caption{The optimal parameters $\Omega_{D0}$ and $f(0)$  for fitting SNe and density perturbations data for model (\ref{EoSBFS}) for various $w_{0}$ and $\epsilon$. The time before future singularity is given also. The coincidence with $\Lambda$CDM model is better for larger values of $\epsilon$ and smaller values of $w_{0}$.}
\end{table}

\begin{figure}
  \includegraphics[scale=1]{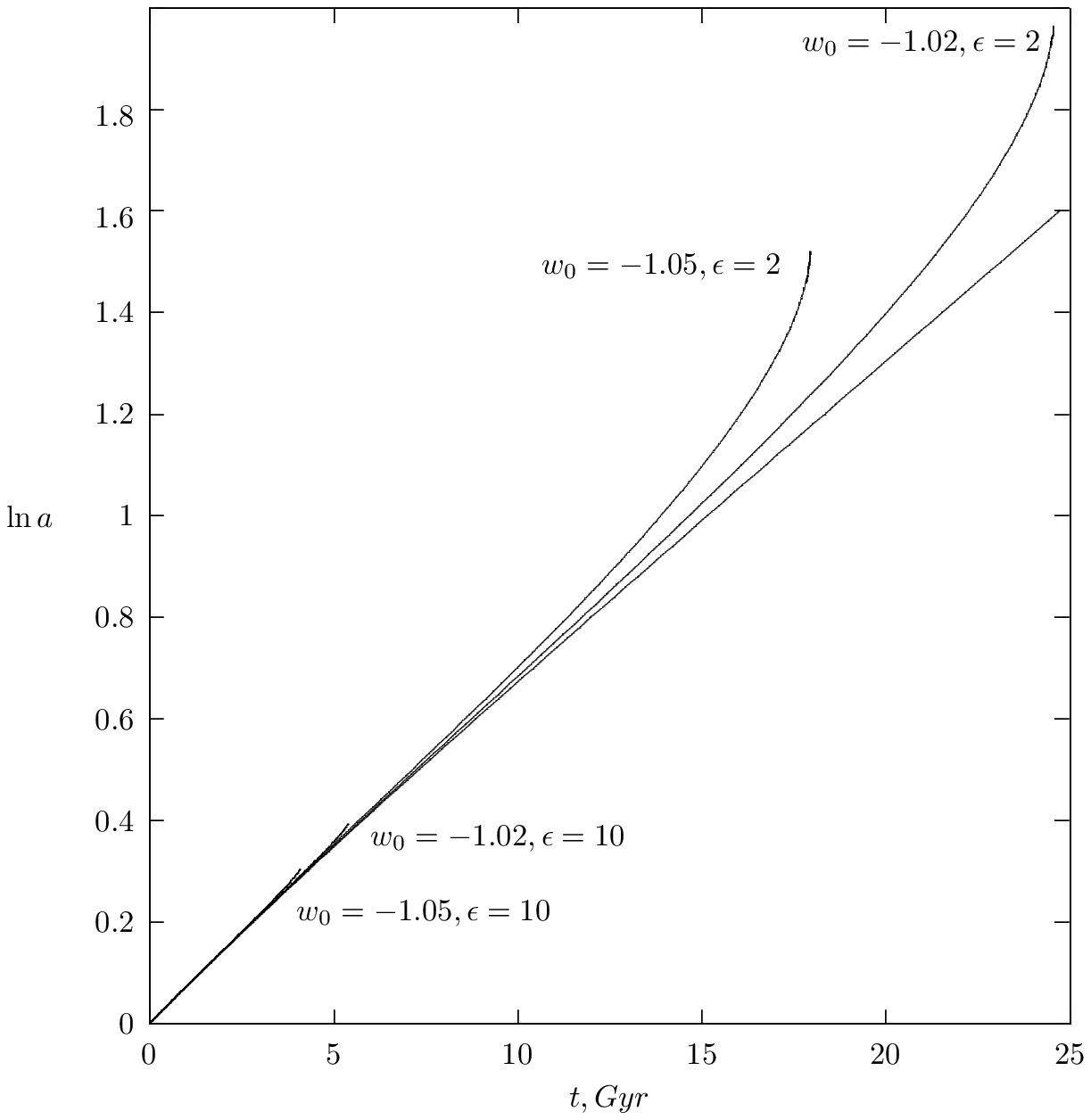}\\
  \caption{The dependence of scale factor from time for DE model (\ref{EoSBFS}) with various parameters choice. For comparison the same dependence for $\Lambda$CDM cosmology (straight line) is given.}
\end{figure}

\textbf{The future evolution of dark energy and matter density perturbations.} To conclude this section we  consider the question about evolution of matter density perturbations in future. In such a case one should account the dark energy density perturbations. The complete system of  cosmological perturbations equations can be found for example in refs.\cite{Perturb}, \cite{Perturb-2}. In the system of units in which $8\pi G=c=1$ and for the case when effective sound speed for dark energy is equal to 1 these equations can be written in the following form
\begin{equation}\label{EQDP}
\delta^{'}_{D}+3(1-w_{D})aH\delta_{D}-(1+w_{D})\delta_{m}+(1+w_{D})\left(k+9a^{2}H^{2}\frac{1-c_{a}^{2}}{k}\right)V_{D}=0,
\end{equation}
\begin{equation}
V_{D}^{'}-2aHV_{D}-\frac{k}{1+w_{D}}\delta_{D}=0,
\end{equation}
\begin{equation}\label{EQMDP}
\delta_{m}^{''}+\frac{a^{'}}{a}\delta^{'}_{m}-\frac{1}{2}(\rho_{m}\delta_{m}+(1+3w_{D})\rho_{D}\delta_{D})=0.
\end{equation}
Here $^{'}$ denotes derivative on conformal time $\tau$ ($\tau=\int a^{-1}dt$), $\delta{D}=\delta\rho_{D}/\rho_{D}$, $V_{D}$ is the velocity perturbation of dark energy and $c_{a}^{2}=dp_{D}/d\rho_{D}$ is the adiabatic speed of sound. The $k$ is a wavenumber of the corresponding mode. One can rewrite Eqs. (\ref{EQDP})-(\ref{EQMDP}) using the relation
$$
d\tau=\frac{da}{a^{2}H}.
$$
Thus, we have
\begin{equation}\label{EQDP-1}
\frac{d\delta_{D}}{da}+\frac{3}{a}(1-w_{D})\delta_{D}-(1+w_{D})\frac{d\delta_{m}}{da}+\frac{1}{a^{2}h}(1+w_{D})\left(\tilde{k}+9a^{2}h^{2}
\frac{1-c_{a}^{2}}{\tilde{k}}\right)V_{D}=0,
\end{equation}
\begin{equation}
\frac{dV_{D}}{da}-\frac{2}{a}V_{D}-\frac{1}{h}\frac{\tilde{k}}{1+w_{D}}\delta_{D}=0,
\end{equation}
\begin{equation}\label{EQMDP-1}
\frac{d^{2}\delta_{m}}{da^{2}}+\left(\frac{3}{a}+\frac{dh}{hda}\right)\frac{d\delta_{m}}{da}=\frac{1}{2a^{2}h^{2}}\frac{1}{H^{2}_{0}}\left(\rho_{m}
\delta_{m}+(1+3w_{D})\rho_{D}\delta_{D}\right),
\end{equation}
where $h=H/H_{0}$ is dimensionless Hubble parameter and $\tilde{k}$ is dimensionless wave number measured in the units of $H_{0}/c$. One notes that last equation of this system is equivalent to (\ref{Perturb}) if  the dark energy density perturbations are neglected. However,  the future DE density increases and even its small perturbations may  become important for the analysis of matter density perturbations (see r.h.s. of Eq. (\ref{EQMDP-1})).

For the model (\ref{LR}) the EoS parameter and adiabatic speed of sound are given as functions of scale factor $$
w(a)=-1-\alpha^{2}\left(1+3\alpha^{2}(1-\beta)\ln a\right)^{-1},\quad c_{a}^{2}=-1+\beta(1+w(a)).
$$
For current value of the scale factor we put simply $a_{0}=1$. Initial conditions for integration are given for example in ref.\cite{Perturb}. The main results of numerical analysis of system (\ref{EQDP-1})-(\ref{EQMDP-1}) are the following. For DE model (\ref{LR}) which evolves to Big Rip at acceptable values of parameters (see Table 3) at small scales (large wavenumbers) the amplitudes of matter density perturbations increase insignificantly in comparison with $\delta_{m0}$ and remain constant. The same picture takes place in $\Lambda$CDM cosmology. However, for the standard cosmological model all modes of matter density perturbations evolve in the same way. In the case under consideration for sufficiently large scales ($>2000$ Mpc) the decay of matter perturbations occurs. This decay occurs faster for perturbations with larger scales (see Fig. 4). The value of dark energy perturbations grows very quickly before $a\sim 10$ but then increases very slowly. If $\beta>1$ (type III singularity) the picture looks the same for matter density perturbations but character of dark energy perturbations evolution changes. We observe the approximately linear growth of $\delta_{D}$ as function of $\ln a$ in the interval $1<a<10$ but then $\delta_{m}$ increases faster according to the law $\delta_{D}\sim (\ln (a_{f}/a))^{-1}$. Shortly before the singularity the perturbations become large. For example for $\beta=3$, $\alpha^{2}=0.03$ the moment of singularity corresponds to $a_{f}=258.67$ ($t_{f}=71.7$ Gyr) and $\delta_{D}\sim 1$ for $k=0.0005$ Mpc$^{-1}$ at $a\sim 235$ ($t=70.4$ Gyr). It is interesting to note that growth of density contrast occurs at sufficiently small background density $\rho_{D}<10\rho_{D0}$. The sharp growth of density perturbations occurs before possible disintegration of such bound structures as Solar System or Milky Way due to enormous acceleration of universe. One can assume that like the formation of large-scale structure due to matter density perturbations in the early universe this growth can lead to the formation of separate domains with various values of dark energy density (``dark large-scale structure''). The cosmological dynamics within these ''sub-universes'' is defined by the background dark energy density. In fact, we cannot speak about the uniform evolution of such universe. It is possible that in some domains the perturbations lead to the decrease of DE density and induce the de-phantomization (one can assume for example that EoS in the form (\ref{LR}) is valid only for $\rho_{D}>\rho^{*}_{D}$ but if $\rho<\rho^{*}_{D}$ the phase transition occurs such that $w>-1$). Yet, in another domains the perturbations of DE energy-density lead to singularity faster than it would be expected without consideration of density perturbations evolution. In a sense this picture maybe similar to chaotic inflation\cite{Linde} reversed in time. The homogeneous universe ends its existence in ``chaotic'' state. Of course, this is rather speculative possibility.

\begin{figure}
  \includegraphics[scale=1]{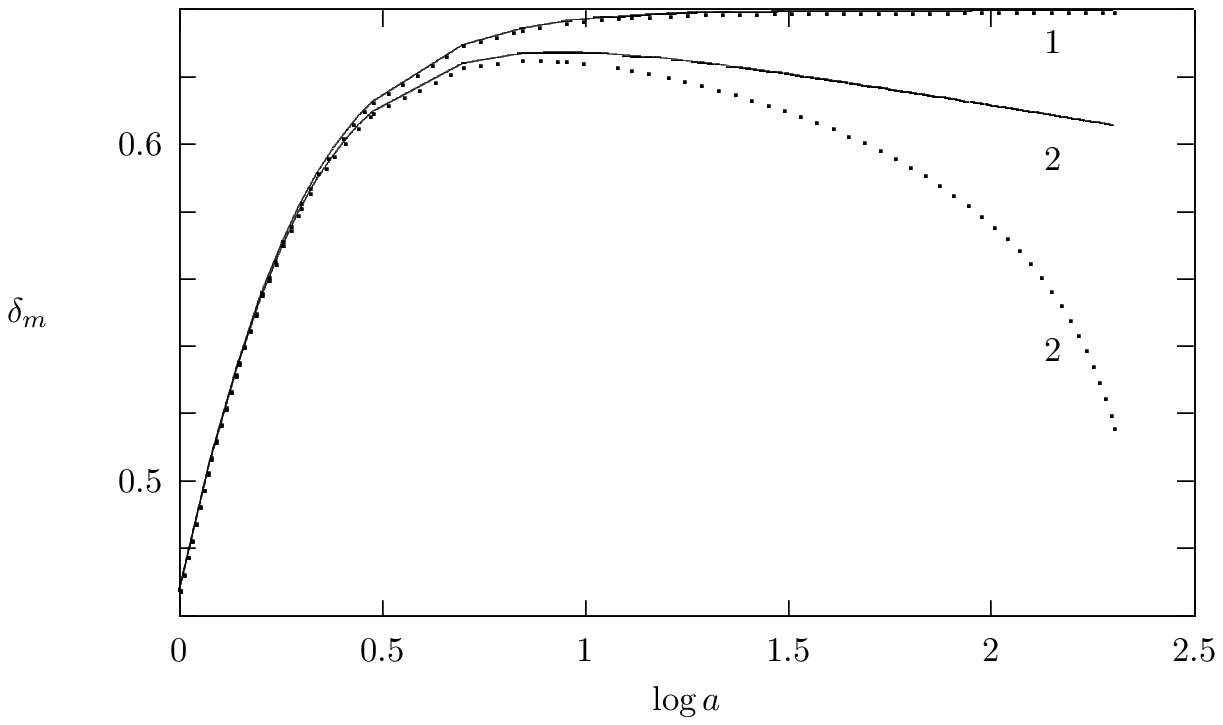}\\
  \includegraphics[scale=1]{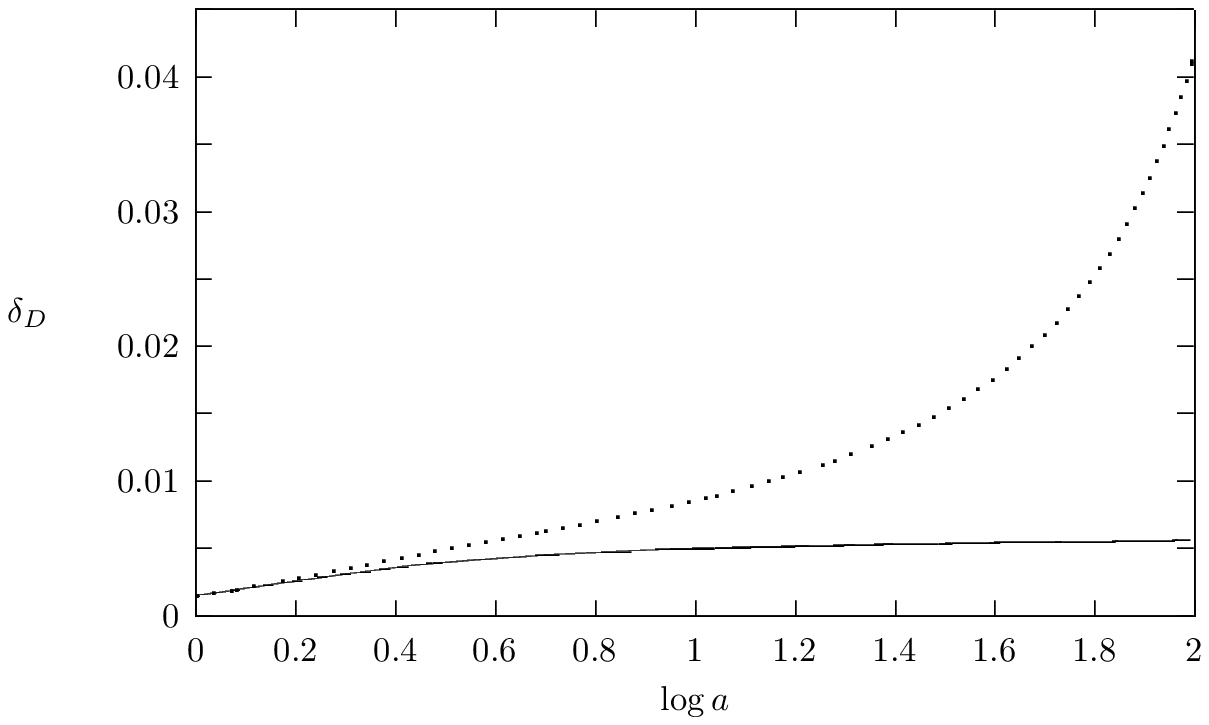}\\
  \caption{The evolution of matter density perturbations (top) and DE energy-density perturbations (bottom) in the model (\ref{LR}) for the cases with Big Rip (bottom lines) and type III singularity (dotted lines). For DE perturbations the evolution of the mode with $k=0.0005$ Mpc$^{-1}$ is depicted. For matter density perturbations the evolution of modes with $k=0.05$ Mpc$^{-1}$ (1) and $k=0.0005$ Mpc$^{-1}$ (2) is shown. The model parameters are chosen as $\alpha^{2}=0.03$ and $\beta=0.95$ and $\beta=3$.}
\end{figure}

\section{DE model II: Pseudo-Rip and type II singularity}

Let us consider another DE model with EoS
\be\label{EOSDS}
g(\rho)=\alpha^{2}\rho_{D0}\left(1-\frac{\rho_{D}}{\rho_{f}}\right)^{\beta}, \quad 0<\rho_{D}<\rho_{f},\quad \beta \neq 0.
\ee
Such dark fluid leads to the following variants of evolution:

(i) DE energy-density asymptotically tends to $\rho_{f}$ if $\beta>1/2$. Therefore, the universe expands  according to de Sitter law at $t\rightarrow\infty$ (Pseudo-Rip).

(ii)DE energy-density reaches $\rho_{f}$ for $t_{f}<\infty$ if $0<\beta\leq 1/2$.

(iii) type II singularity occurs if $\beta<0$. The second derivative of scale factor diverges while first derivative remains finite.

Let us consider two simplest cases: $\beta=1$, $-1$.

(a) $\beta=1$. For EoS (\ref{EOSDS}) the dark energy density as function of redshift is
\be\label{rhoEOSDS}
\rho_{D}(z)=\rho_{D0}\Delta^{-1}\left(1-(1-\Delta)(1+z)^{3\alpha^{2}\Delta}\right), \Delta=\rho_{D0}/\rho_{f}.
\ee
As in previous section we consider only SNe data first. For various $\Delta$ one can find optimal values of parameters $\alpha$ and $\Omega_{D0}$. Results are given in table V.

\begin{table}
\label{Table6}
\begin{centering}

\textbf{DE models with Pseudo-Rip: analysis of SNe data}

\begin{tabular}{|c|c|c|c|c|c|c|c|c|c|c|}
  \hline
& \multicolumn{2}{c}{$\Delta=0.1$}\vline & \multicolumn{2}{c}{$\Delta=0.2$}\vline & \multicolumn{2}{c}{$\Delta=0.5$}\vline & \multicolumn{2}{c}{$\Delta=0.8$}\vline & \multicolumn{2}{c}{$\Delta=0.95$}\vline \\
  \hline
  $\alpha^{2}$ & $\Omega_{D0}$ & $\chi^{2}_{SN}$ & $\Omega_{D0}$ & $\chi^{2}_{SN}$ & $\Omega_{D0}$ & $\chi^{2}_{SN}$ & $\Omega_{D0}$ & $\chi^{2}_{SN}$ & $\Omega_{D0}$ & $\chi^{2}_{SN}$\\

\hline
  0.1 & 0.68 & 347.30 & 0.68 & 347.25 & 0.69 & 347.15 & 0.71 & 347.04 & 0.71 & 347.00\\
  0.2 & 0.64 & 348.17 & 0.65 & 347.85 & 0.67 & 347.38 & 0.70 & 347.07 & 0.71 & 346.96\\
  0.3 & 0.61 & 349.17 & 0.62 & 348.72 & 0.65 & 347.71 & 0.69 & 347.12 & 0.71 & 346.99\\
  0.4 & 0.59 & 350.54 & 0.60 & 349.83 & 0.63 & 348.10 & 0.68 & 347.20 & 0.71 & 347.09\\
  0.5 & 0.56 & 351.66 & 0.57 & 350.69 & 0.61 & 348.50 & 0.67 & 347.28 & 0.70 & 347.00\\
  \hline
\end{tabular}
\caption{The optimal value of parameter $\Omega_{D0}$ and corresponding value of $\chi^{2}_{SN}$ for some $\alpha^{2}$ and $\Delta$.}
\end{centering}

\end{table}

The analysis of density perturbations data shows that  only in the narrow range of $\alpha^{2}$ these data can be described with 95\% C.L. (see table VI). This fact simply means that current value of EoS parameter
$$
w_{0}=-1-\alpha^{2}(1-\Delta)
$$
only insignificantly deviates from $-1$.

Note that same analysis can be performed for quintessence DE model
 with asymptotic de Sitter evolution. The parameter $\Delta$ in this case is larger than $1$. The Eq. (\ref{rhoEOSDS}) remains correct. The analysis shows that for $\Delta\sim1$ and small $\alpha^{2}$ the model describes observational data with good precision.

\begin{table}
\label{Table7}
\begin{centering}

\textbf{DE models with Pseudo-Rip: analysis of SNe+DP data}

\begin{tabular}{|c|c|c|c|c|c|}
\hline
$\Delta$ & $0.1$ & $0.2$ & $0.5$ & $0.8$ & $0.95$  \\
\hline
$\alpha^{2}_{max}$ & 0.03 & 0.03 & 0.06 & 0.13 & 0.45 \\
\hline
$\Omega_{D0}$ & \multicolumn{5}{c}{0.71}\vline \\
\hline
$f(0)$ & \multicolumn{5}{c}{0.514}\vline \\
\hline
$\chi^{2}_{SN} $ & 347.19 & 347.11 & 347.30 & 347.19 & 346.98 \\
\hline
$A$ &  \multicolumn{5}{c}{0.482}\vline  \\
\hline
\end{tabular}
\caption{The maximal value of $\alpha^{2}$ for various $\Delta$ and $f(0)$ at which the model (\ref{EOSDS}) describes the density perturbations data with 95\% C.L. The optimal value of $\Omega_{D0}$ for fitting SNe data, $\chi^{2}_{SN}$ and BAO parameter are also given.}
\end{centering}
\end{table}

(b) $\beta=-1$. In this case for DE energy-density as function of redshift we can derive the relation
\be
\rho_{D}(z)=\rho_{D0}\Delta^{-1}\left\{1-((1-\Delta)^{2}+6\alpha^{2}\Delta\ln(z+1))^{1/2}\right\}
\ee
The time left for future singularity can be found from (\ref{trho}) by integrating from $\rho_{D0}$ to $\rho_{f}$. We have
\be
t_{f}-t_{0}=\frac{2}{3\alpha^{2}\Omega^{1/2}_{D0}\Delta^{1/2}}\left(\frac{2}{3}-\Delta^{1/2}\left(1-\frac{1}{3}\Delta\right)\right)
\ee
Next, as in the previous cases we calculated the optimal value of $\Omega_{D0}$ for fitting SNe fata for given $\Delta$ and $\alpha^{2}$ (table VII). One sees that only at sufficiently large $\alpha^{2}$ and $\Delta$ our model declines from observational data significantly. For $\Delta\rightarrow 1$ the model (\ref{EOSDS}) fits the observational data only at $\alpha^{2}\rightarrow 0$.

For current value of EoS parameter in the model with type II singularity we have
$$
w_{0}=-1-\frac{\alpha^{2}}{1-\Delta}.
$$
Again, $w_{0}$ is close to $-1$.

\begin{table}
\label{Table8}
\begin{centering}

\textbf{DE models with type II singularity: analysis of SNe data}

\begin{tabular}{|c|c|c|c|c|c|c|}
  \hline
& \multicolumn{3}{c}{$\Delta=0.2$}\vline & \multicolumn{3}{c}{$\Delta=0.5$}\vline \\
  \hline
  $\alpha^{2}$ & $\Omega_{D0}$ & $t_{f}$, Gyr & $\chi^{2}_{SN}$ & $\Omega_{D0}$ & $t_{f}$, Gyr & $\chi^{2}_{SN}$ \\

\hline
  0.1  & 0.67 & 27.6 & 347.67 & 0.64 & 8.7 & 348.64 \\
  0.2  & 0.62 & 14.2 & 349.02 & 0.59 & 4.6 & 351.75  \\
  0.3  & 0.59 & 9.8 & 351.04 & 0.54 & 3.2 & 356.14  \\
  0.4  & 0.55 & 7.6 & 353.18 & 0.51 & 2.5 & 360.67  \\
  0.5  & 0.52 & 6.2 & 355.65 & 0.48 & 2.0 & 365.51  \\
  \hline
\end{tabular}

\caption{The optimal value of parameter $\Omega_{D0}$, time before singularity and corresponding value of $\chi^{2}_{SN}$ for some $\alpha^{2}$ and $\Delta$.}
\end{centering}
\end{table}

The combined analysis of SNe data and data for density perturbations gives the results similar to previous cases. The deviation from $\Lambda$CDM model is sufficiently small for fitting observational data although these data in principle do not prevent the possibility of future singularity within $\sim100$ Gyr (table VIII).

\begin{table}
\label{Table9}
\begin{centering}

\textbf{DE models with type II singularity: analysis of SNe+DP data}

\begin{tabular}{|c|c|c|}
\hline
$\Delta$ & $0.2$ & $0.5$  \\
\hline
$\alpha^{2}_{max}$ & 0.02 & 0.012 \\
\hline
$\Omega_{D0}$ & \multicolumn{2}{c}{0.71}\vline \\
\hline
$t_{fmin}$, Gyr & 134.1 & 69.4 \\
\hline
$f(0)$ & \multicolumn{2}{c}{0.513}\vline \\
\hline
$\chi^{2}_{SN} $ & 347.13 & 347.10 \\
\hline
$A$ & \multicolumn{2}{c}{0.482}\vline\\
\hline
\end{tabular}

\caption{The maximal value of $\alpha^{2}$ for various $\Delta$ and $f(0)$ at which the model (\ref{EOSDS}) describes the density perturbations data with 2$\sigma$ CL. The optimal value of $\Omega_{D0}$ for fitting SNe data, $\chi^{2}_{SN}$ and BAO parameter are also given.}
\end{centering}

\end{table}

The analysis of the evolution of density perturbations can be performed as at the end of previous section. Let us briefly recall the main results. For Pseudo-Rip the matter density perturbations evolve in the same way as in $\Lambda$CDM model. For dark energy density perturbations we have rapid growth before $a\sim 10$ and then slow decay. For models with type II singularity the picture coincides with that  of (\ref{LR}) with type III singularity. Hence, DE models mimicking $\Lambda$CDM and fitting current observational bounds may show different exotic behaviour in the future: finite-time singularities, disintegration of bound structures(Little Rip or Pseudo-Rip cosmologies) or decay of cosmological perturbations.

\section{Conclusion}

In summary,
we confronted number of DE models  mimicking $\Lambda$CDM epoch with current EoS parameter being very close to $-1$  with combined observational data: SNe data, baryon acoustic oscillations data
and DE and matter energy-density perturbations. It is explicitly demonstrated that  there exists sufficiently wide region of parameters for each of DE model under discussion where these theories are not less viable than the standard $\Lambda$CDM model.
On the same time, DE models under consideration show qualitatively different future behaviour with Big Rip, type II and type III future singularities, Little Rip, Pseudo-Rip or Quasi-Rip evolution.
Nevertheless, current observational data cannot determine whether or not the universe will end in a future singularity.

It should be noted that the observational data for density perturbations are more sensitive to the deviation from standard
cosmological model in comparison with SNe data. Moreover, the account of density perturbations in Rip cosmology indicates to sharp growth of density at sufficiently small background density still before the possible disintegration of bound objects. This growth from our viewpoint can lead to possibility that future universe may split in the number of separate regions so that it becomes chaotic and never reaches the Rip singularity. Further consequences of above effect will be considered elsewhere.

\section*{Acknowledgments}
We are grateful to M. Trodden, M. Sasaki and R. Scherrer for useful discussions.
This work has been partly
supported by MICINN (Spain), project FIS2010-15640
and by AGAUR (Generalitat de Catalunya), contract 2009SGR-994.

\section*{Appendix: DE model with dephantomization: Quasi-Rip}

In this Appendix we briefly discuss the possibility of so-called ``Quasi-Rip'' scenario considered  in ref.\cite{Wei-0}. Let us consider the case when DE pressure
 depends from its energy-density as follows:
\be\label{QR}
p_{D}=\left\{\begin{array}{l|l}
-\rho_{D}-\frac{2}{3}\alpha^{2}\rho_{D}\left(\ln\frac{\rho_{m}}{\rho_{D}}\right)^{1/2},\quad a<a_{T},\\
-\rho_{D}+\frac{2}{3}\alpha^{2}\rho_{D}\left(\ln\frac{\rho_{m}}{\rho_{D}}\right)^{1/2},\quad a>a_{T},
\end{array}
\right.
\ee
where $\rho_{D}\leq\rho_{m}$ and $\ln a_{T}=\alpha^{-2}\left(\ln(\rho_{m}/\rho_{D0})\right)^{1/2}$ is a value of scale factor at which the dephantomization occurs. In this moment the value of DE energy-density reaches the maximal value $\rho_{m}$. For small $\alpha^{2}$ the universe acceleration  is maximal for $\rho_{D}\approx \rho_{m}\exp(-\alpha^{2}/2)$. The pressure as function of  energy-density is depicted on Fig. 5.

\begin{figure}
  \includegraphics[scale=1]{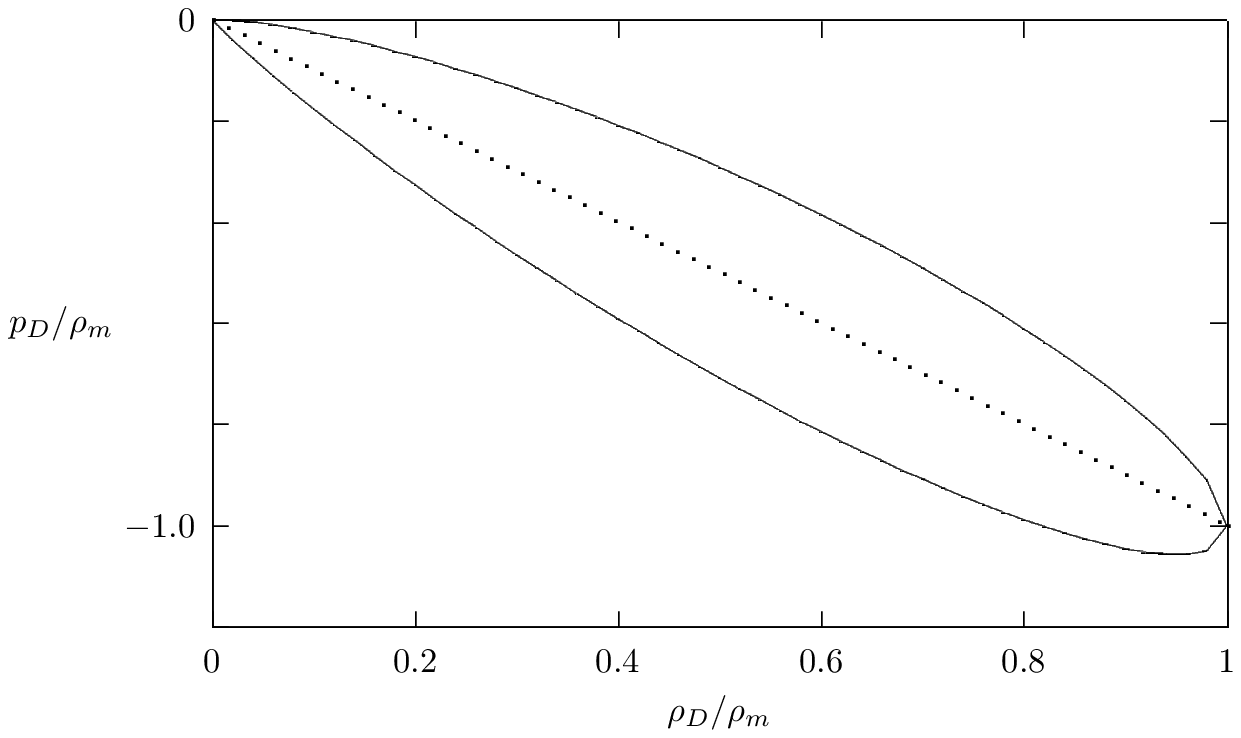}\\
  \caption{The pressure as function of  energy-density for model (\ref{QR}). The dotted line corresponds to vacuum energy with $w=-1$. }
\end{figure}

The dependence of DE energy-density from redshift $z$ for EoS (\ref{QR}) is
\be
\rho_{D}(z)=\rho_{m}\exp\left(-\alpha^{4}(\ln a_{T}+\ln (z+1))^{2}\right)
\ee
One can rewrite this equation as
\be\label{rhoQR}
\rho_{D}=\rho_{D0}\exp\left(-\gamma\ln^{2}(z+1)-\beta\ln(z+1)\right), \quad \gamma=\alpha^{4},\quad \beta=2\alpha^{4}\ln a_{T}.
\ee
The Eq. (\ref{rhoQR}) coincides with the equation considered in ref. \cite{Wei-0}. As is demonstrated in above
 work this model fits modern SNe data in the large interval of free parameters. One can show that the model (\ref{QR}) is consistent with matter density perturbations data also. Hence, the Quasi-Rip model maybe consistent with current observational data.

The specific feature of Eq. (\ref{QR}) is that scale factor is double-valued function of the energy-density. The branch point corresponds to dephantomization: the pressure in this point is defined uniquely.
It is obvious that the described scheme can be generalized. For equations of state with branch points the evolution of universe contains (de)phantomization and  Quasi-Rip epochs.

\end{document}